\documentclass[12pt, a4paper]{article}
\pdfoutput=1
\usepackage[latin1]{inputenc}
\usepackage{amsmath}
\usepackage{amsfonts}
\usepackage{amssymb}
\usepackage{latexsym}
\usepackage{mathrsfs}
\usepackage{graphicx}
\usepackage{color}
\usepackage{twistor}
\usepackage[all]{xy}
\usepackage[colorlinks=true, linkcolor=black, citecolor=black, urlcolor=black, linktocpage=true]{hyperref}

\renewcommand{\d}{\mathrm{d}}

\begin{document}

\title{\Large General relativity as a two-dimensional CFT\thanks{Honorable Mention in the 2015 Gravity Research Foundation Essay Competition}}

\author{\normalsize Tim Adamo \\ \small \textit{Department of Applied Mathematics \& Theoretical Physics} \\ \small \textit{University of Cambridge} \\ \small \textit{Wilberforce Road} \\ \small \textit{Cambridge, CB3 0WA, U.K.} \\ \small \texttt{t.adamo@damtp.cam.ac.uk}}

\date{}

\maketitle

\abstract{The tree-level scattering amplitudes of general relativity encode the full non-linearity of the Einstein field equations. Yet remarkably compact expressions for these amplitudes have been found which seem unrelated to a perturbative expansion of the Einstein-Hilbert action. This suggests an entirely different description of GR which makes this on-shell simplicity manifest. Taking our cue from the tree-level amplitudes, we discuss how such a description can be found. The result is a formulation of GR in terms of a solvable two-dimensional CFT, with the Einstein equations emerging as quantum consistency conditions.}

\vfill

\pagebreak


As a quantum field theory, general relativity (GR) is non-renormalizable and therefore fails to make sense at high energies. Nevertheless, it provides an important litmus test which \emph{any} purported theory of quantum gravity must pass by reducing to GR in the semi-classical, or low-energy, limit. In other words, the low-energy observables of any theory of quantum gravity should be equal to those of GR. In an asymptotically flat space-time, tree-level scattering amplitudes are natural examples of such semi-classical observables: they provide `theoretical data' against which a theory of quantum gravity can be checked. Due to diffeomorphism invariance, these scattering amplitudes are not just natural, they are the \emph{only} semi-classical observables~\cite{DeWitt:1967ub}.

The tree-level S-matrix also acts as a test for the complexity of classical field theories, and GR is no exception. While its computation is usually presented in terms of Feynman diagrams, it can also be understood -- and in principle, computed -- from a purely variational perspective. Given a classical action functional, the tree-level S-matrix is constructed by perturbatively expanding the action with respect to linearized states propagating on a non-linear background (usually chosen to be trivial). Feynman rules, derived from the action, give an intuitive method for organizing this perturbative expansion. While a single scattering amplitude probes the action by linearized fields, knowing the full tree-level S-matrix of a theory is equivalent to knowing its non-linear equations of motion. 

For general relativity, the non-linear equations in question are the vacuum Einstein equations:
\be\label{EFEs}
R = 0 = R_{\mu\nu}\,,
\ee
where $R_{\mu\nu}$ is the Ricci tensor and $R$ the Ricci scalar. These are the field equations of the Einstein-Hilbert action
\be\label{EH}
S[g]=\frac{1}{16\pi G_{\mathrm{N}}}\int_{M}\d^{d}x\,\sqrt{-g}\,R\,,
\ee
where $d$ is the number of space-time dimensions. Unfortunately, the perturbative expansion of the Einstein-Hilbert action, as operationalized by Feynman diagrams, is a nightmare, containing an \emph{infinite} number of interaction vertices~\cite{DeWitt:1967ub,DeWitt:1967uc,'tHooft:1974bx}. Consequently, it seems that the classical scattering amplitudes of GR will be a complicated, unenlightening mess. 

Yet over half a century of research has uncovered surprises undermining the complexity implied by \eqref{EH}. From the realization that terms in the Feynman expansion of the four-point amplitude (over \emph{five hundred} in number) sum to a simple expression (\textit{c.f.}, \cite{DeWittMorette:1971zz,DeWitt:2008vw}), to loop-level computations raising the possibility that maximal supergravity could be a finite quantum field theory in four dimensions~\cite{Bern:2011qn}, it is clear that scattering amplitudes in gravity have myriad properties which could never be guessed from \eqref{EH}.\footnote{Dramatic progress in the study of scattering amplitudes -- especially in the last decade -- has been made for a wide array of theories, including gauge theory and string theory (\textit{c.f.}, \cite{Elvang:2013cua}).} Regarding the tree-level amplitudes of GR, one particular development stands out: an expression for the \emph{entire} tree-level (Minkowski space) S-matrix in any dimension.

Remarkably, this expression -- found by Cachazo, He, and Yuan (CHY)~\cite{Cachazo:2013hca} -- has no origin in the Feynman rules of the Einstein-Hilbert action. Indeed, rather than a sum over Feynman diagrams, the CHY formula is given by an integral over the moduli space of a Riemann sphere, $\Sigma$, with marked points, $\{z_{1},\ldots,z_{n}\}$. The locations of the $z_i$s are fixed in terms of the null momenta of the external states, $\{k_{1},\ldots,k_{n}\}$, by a set of $n-3$ constraints, known as \emph{scattering equations}: 
\be\label{SEqs}
\mathcal{S}_{i}=\sum_{j\neq i}\frac{k_{i}\cdot k_{j}}{z_{i}-z_{j}}=0\,, \quad i=4,\ldots,n\,.
\ee
Schematically, the formula for the $n$-point tree-level amplitude is
\be\label{CHY}
\cM_{n}=\int \d\mu_{n}\,\prod_{i}\,^{\prime}\,\delta\left(\mathcal{S}_i\right)\,\mathcal{I}_{n}\,,
\ee
where $\d\mu_{n}$, $\prod_{i}^{\prime}$ denote M\"obius-invariant measures and products on the moduli space, and $\mathcal{I}_{n}$ is an integrand depending on the $\{z_{i}\}$ and kinematic data -- see~\cite{Cachazo:2013hca} for details.

We claim that there is a fundamental statement about GR implied by the disconnect between \eqref{CHY} and a perturbative expansion of \eqref{EH}. That is, the unexpected simplicity of \eqref{CHY} hints at an alternative formulation of GR making this structure manifest. As we will see, not only does this alternative exist, but it takes a surprisingly elegant form~\cite{Adamo:2014wea}.

\medskip

The structure of \eqref{CHY} shares several similarities with the tree-level S-matrix of string theory, including moduli space integrals, compact amplitude expressions, and manifestation of low-energy behavior through degenerations of the underlying Riemann sphere (\textit{c.f.}, \cite{Mafra:2011nv,Witten:2012bh}).\footnote{Surprisingly, the scattering equations also appear in the high-energy regime of string theory~\cite{Gross:1987ar}.}  However, string theory depends on an additional parameter: the string length $\sqrt{\alpha^{\prime}}$. Taking $\alpha^{\prime}\rightarrow 0$ gives field theory amplitudes~\cite{Scherk:1974ca}, so \eqref{CHY} seems -- heuristically, at least -- to arise from some ``infinite tension'' limit of string theory.

Indeed, soon after the appearance of \eqref{CHY}, it was realized that the CHY formulae were equal to correlation functions of a certain chiral, first-order conformal field theory (CFT) on the Riemann sphere~\cite{Mason:2013sva}. The structure of this theory is akin to a complexification of the worldline action for supersymmetric quantum mechanics (\textit{c.f.}, \cite{Blau:1992pm}); there is no $\alpha^{\prime}$ parameter and the scattering equations emerge from a gauge-fixing procedure~\cite{Adamo:2013tsa}. Since it computes the full tree-level S-matrix of GR on Minkowski space, formulating this theory for a general curved space-time should give a non-linear description of GR.  

The formulation of this curved-space model is relatively straightforward at the classical level. Consider a space-time $M$ of dimension $d$ equipped with a metric $g_{\mu\nu}$, and a closed Riemann surface $\Sigma$ (the ``worldsheet'') with a complex structure represented by the anti-holomorphic Dolbeault operator $\dbar$. The model is defined by the two-dimensional action functional~\cite{Adamo:2014wea}:
\be\label{WSA1}
S=\frac{1}{2\pi}\int_{\Sigma} P_{\mu}\,\dbar X^{\mu}+\bar{\psi}_{\mu}\,\bar{D}\psi^{\mu}\,.
\ee
Here, $X^{\mu}$ are coordinates on $d$-dimensional space-time which are scalars on $\Sigma$, while $P_{\mu}$ is a space-time covector taking values in the space of worldsheet $(1,0)$-forms. The pair $\bar{\psi}_{\mu},\,\psi^{\nu}$ transform as space-time covectors and vectors, respectively; both are worldsheet spinors with fermionic statistics. The operator $\bar{D}$ is the anti-holomorphic covariant derivative, pulled back to the worldsheet:
\begin{equation*}
 \bar{D}\psi^{\mu}=\dbar\psi^{\mu}+\Gamma^{\mu}_{\nu\rho}\psi^{\nu}\dbar X^{\rho}\,,
\end{equation*}
where $\Gamma^{\mu}_{\nu\rho}$ are the Christoffel symbols of the metric $g_{\mu\nu}$ on $M$.

Since the action \eqref{WSA1} depends only on the complex structure of $\Sigma$, it is a two-dimensional CFT. As a classical action, it is also invariant under diffeomorphisms of $M$. However, this invariance is far from obvious at the quantum level. That is, the path integral defined by \eqref{WSA1} may fail to be invariant under space-time diffeomorphisms.

Under a diffeomorphism on $M$, the bosonic portion of \eqref{WSA1} must be shifted by a level-one Wess-Zumino-Witten (WZW) term to remove potentially anomalous two-point functions~\cite{Nekrasov:2005wg,Witten:2005px}. For the fermions, this diffeomorphism corresponds to a chiral rotation; while the classical action is not modified, the path integral measure is corrected by a phase. This phase is precisely the same WZW action produced by the bosons, with the opposite sign~\cite{Polyakov:1984et}. Hence, the path integral for the theory \emph{is} invariant under a diffeomorphism and thus makes sense with respect to smooth coordinate transformations on $M$.

\medskip

So \eqref{WSA1} defines a two-dimensional CFT which is diffeomorphism invariant at both the classical and quantum level. Additionally, this theory has the remarkable property that it becomes \emph{free} after a field redefinition
\be\label{FR}
\Pi_{\mu}= P_{\mu}+\Gamma_{\mu\nu}^{\rho}\bar{\psi}_{\rho}\psi^{\nu}\,,
\ee
giving a purely kinetic action:
\be\label{WSA2}
S=\frac{1}{2\pi}\int_{\Sigma}\Pi_{\mu}\,\dbar X^{\mu}+\bar{\psi}_{\mu}\,\dbar\psi^{\mu}\,.
\ee
This means that the operator product expansions (OPEs) between worldsheet fields are free, so any quantity in the CFT can be computed exactly, without recourse to perturbation theory.

In addition to conformal symmetry on $\Sigma$, the action \eqref{WSA2} has three additional symmetries: a Hamiltonian (generated by a Laplacian on $M$), and two fermionic symmetries (analogous to chiral supersymmetries). Gauge-fixing these symmetries leads to potential anomalies in the model. These are a conformal anomaly and an anomaly in a certain worldsheet current algebra related to the fermionic and Hamiltonian symmetries~\cite{Adamo:2014wea}. While the former is eliminated in the critical space-time dimension $d=10$ and doesn't affect correlation functions on the Riemann sphere, the current algebra anomaly is more subtle.

Though the model is sensible with respect to diffeomorphisms of space-time, its symmetries do not behave correctly under diffeomorphisms of the \emph{worldsheet}. Crucially, this is remedied by modifying \eqref{WSA2} in a way that does not affect the conformal anomaly or the theory's OPEs. With these free OPEs, the current algebra anomaly can be computed \emph{exactly}, and is given by the Einstein equations \eqref{EFEs}.\footnote{A $B$-field and dilaton are easily incorporated into the model, leading to the field equations for the NS-NS sector of type II supergravity.} In other words, the field equations emerge as the quantum consistency conditions of this two-dimensional CFT~\cite{Adamo:2014wea}. This should be contrasted with the analogous calculation for string theory, where anomalies can only be computed perturbatively in $\alpha^{\prime}$ via a complicated background field expansion~\cite{Callan:1985ia}.

\medskip

So GR can be described by a solvable two-dimensional CFT with free OPEs; there is no need to make reference to a space-time action principle at all! In the context of Minkowski space, this statement manifests itself in the structure of \eqref{CHY}, given by a single correlation function on the Riemann sphere as opposed to a morass of Feynman diagrams coming from the Einstein-Hilbert action~\cite{Mason:2013sva}. Beyond the classical level, correlation functions on higher-genus Riemann surfaces appear to compute compact expressions for the loop integrands of (super-)gravity~\cite{Adamo:2013tsa,Casali:2014hfa}.

More generally, this reformulation implies that the calculation of \emph{any} semi-classical observable in GR is equivalent to computing a correlation function in a two-dimensional CFT with free OPEs. While the consequences of this statement beyond the confines of Minkowski space have yet to be explored, it seems reasonable to expect it to shed new light on the computation of S-matrices in non-flat space-times -- a subject hitherto hamstrung by the complexity of the Einstein-Hilbert action's perturbation theory. In any case, the fact that our basic theory of gravitation can be formulated in this way, using tools which seem so disjoint from standard space-time-based approaches, indicates that even a century after its discovery, general relativity remains an intriguing field of research. 

\subsubsection*{Acknowledgments}

This work is supported by a Title A Research Fellowship at St. John's College, Cambridge. The results reported in this essay were found in collaboration with Eduardo Casali and David Skinner~\cite{Adamo:2014wea}.

\bibliography{GRF}

\providecommand{\href}[2]{#2}\begingroup\raggedright\begin{thebibliography}{10}

\bibitem{DeWitt:1967ub}
B.~S. DeWitt, {\it {Quantum Theory of Gravity. 2. The Manifestly Covariant
  Theory}},  {\em Phys.Rev.} {\bf 162} (1967) 1195--1239.

\bibitem{DeWitt:1967uc}
B.~S. DeWitt, {\it {Quantum Theory of Gravity. 3. Applications of the Covariant
  Theory}},  {\em Phys.Rev.} {\bf 162} (1967) 1239--1256.

\bibitem{'tHooft:1974bx}
G.~'t~Hooft and M.~Veltman, {\it {One loop divergencies in the theory of
  gravitation}},  {\em Annales Poincare Phys.Theor.} {\bf A20} (1974) 69--94.

\bibitem{DeWittMorette:1971zz}
C.~DeWitt-Morette and W.~Wesley, {\it {Quantum falling charges}},  {\em
  Gen.Rel.Grav.} {\bf 2} (1971) 235--245.

\bibitem{DeWitt:2008vw}
B.~DeWitt, {\it {Quantum Gravity Yesterday and Today}},  {\em Gen.Rel.Grav.}
  {\bf 41} (2009) 413--419, [\href{http://arxiv.org/abs/0805.2935}{{\tt
  arXiv:0805.2935}}].

\bibitem{Bern:2011qn}
Z.~Bern, J.~J. Carrasco, L.~J. Dixon, H.~Johansson, and R.~Roiban, {\it
  {Amplitudes and Ultraviolet Behavior of N = 8 Supergravity}},  {\em
  Fortsch.Phys.} {\bf 59} (2011) 561--578,
  [\href{http://arxiv.org/abs/1103.1848}{{\tt arXiv:1103.1848}}].

\bibitem{Elvang:2013cua}
H.~Elvang and Y.-t. Huang, {\it {Scattering Amplitudes}},
  \href{http://arxiv.org/abs/1308.1697}{{\tt arXiv:1308.1697}}.

\bibitem{Cachazo:2013hca}
F.~Cachazo, S.~He, and E.~Y. Yuan, {\it {Scattering of Massless Particles in
  Arbitrary Dimensions}},  {\em Phys.Rev.Lett.} {\bf 113} (2014), no.~17
  171601, [\href{http://arxiv.org/abs/1307.2199}{{\tt arXiv:1307.2199}}].

\bibitem{Adamo:2014wea}
T.~Adamo, E.~Casali, and D.~Skinner, {\it {A Worldsheet Theory for
  Supergravity}},  {\em JHEP} {\bf 1502} (2015) 116,
  [\href{http://arxiv.org/abs/1409.5656}{{\tt arXiv:1409.5656}}].

\bibitem{Mafra:2011nv}
C.~R. Mafra, O.~Schlotterer, and S.~Stieberger, {\it {Complete N-Point
  Superstring Disk Amplitude I. Pure Spinor Computation}},  {\em Nucl.Phys.}
  {\bf B873} (2013) 419--460, [\href{http://arxiv.org/abs/1106.2645}{{\tt
  arXiv:1106.2645}}].

\bibitem{Witten:2012bh}
E.~Witten, {\it {Superstring Perturbation Theory Revisited}},
  \href{http://arxiv.org/abs/1209.5461}{{\tt arXiv:1209.5461}}.

\bibitem{Gross:1987ar}
D.~J. Gross and P.~F. Mende, {\it {String Theory Beyond the Planck Scale}},
  {\em Nucl.Phys.} {\bf B303} (1988) 407.

\bibitem{Scherk:1974ca}
J.~Scherk and J.~H. Schwarz, {\it {Dual Models for Nonhadrons}},  {\em
  Nucl.Phys.} {\bf B81} (1974) 118--144.

\bibitem{Mason:2013sva}
L.~Mason and D.~Skinner, {\it {Ambitwistor strings and the scattering
  equations}},  {\em JHEP} {\bf 1407} (2014) 048,
  [\href{http://arxiv.org/abs/1311.2564}{{\tt arXiv:1311.2564}}].

\bibitem{Blau:1992pm}
M.~Blau, {\it {The Mathai-Quillen formalism and topological field theory}},
  {\em J.Geom.Phys.} {\bf 11} (1993) 95--127,
  [\href{http://arxiv.org/abs/hep-th/9203026}{{\tt hep-th/9203026}}].

\bibitem{Adamo:2013tsa}
T.~Adamo, E.~Casali, and D.~Skinner, {\it {Ambitwistor strings and the
  scattering equations at one loop}},  {\em JHEP} {\bf 1404} (2014) 104,
  [\href{http://arxiv.org/abs/1312.3828}{{\tt arXiv:1312.3828}}].

\bibitem{Nekrasov:2005wg}
N.~A. Nekrasov, {\it {Lectures on curved beta-gamma systems, pure spinors, and
  anomalies}},  \href{http://arxiv.org/abs/hep-th/0511008}{{\tt
  hep-th/0511008}}.

\bibitem{Witten:2005px}
E.~Witten, {\it {Two-dimensional models with (0,2) supersymmetry: Perturbative
  aspects}},  {\em Adv.Theor.Math.Phys.} {\bf 11} (2007)
  [\href{http://arxiv.org/abs/hep-th/0504078}{{\tt hep-th/0504078}}].

\bibitem{Polyakov:1984et}
A.~M. Polyakov and P.~Wiegmann, {\it {Goldstone Fields in Two-Dimensions with
  Multivalued Actions}},  {\em Phys.Lett.} {\bf B141} (1984) 223--228.

\bibitem{Callan:1985ia}
C.~G. Callan, E.~Martinec, M.~Perry, and D.~Friedan, {\it {Strings in
  Background Fields}},  {\em Nucl.Phys.} {\bf B262} (1985) 593.

\bibitem{Casali:2014hfa}
E.~Casali and P.~Tourkine, {\it {Infrared behaviour of the one-loop scattering
  equations and supergravity integrands}},  {\em JHEP} {\bf 1504} (2015) 013,
  [\href{http://arxiv.org/abs/1412.3787}{{\tt arXiv:1412.3787}}].

\end{thebibliography}\endgroup
\bibliographystyle{JHEP}

\end{document}